\documentclass[aps,prl,twocolumn,amsmath,amssymb,tightenlines,epsfig,floatfix,superscriptaddress]{revtex4-1}
\usepackage{graphicx}
\usepackage{dcolumn}	
\usepackage{bm}			
\usepackage{amsfonts}
\usepackage{xspace}
\usepackage{color}
\usepackage{epstopdf}
\usepackage{multirow}

\begin{document}

\newcommand{\psihat}{\ensuremath{\hat{\psi}}\xspace}
\newcommand{\psihatd}{\ensuremath{\hat{\psi}^{\dagger}}\xspace}
\newcommand{\ahat}{\ensuremath{\hat{a}}\xspace}
\newcommand{\Ham}{\ensuremath{\mathcal{H}}\xspace}
\newcommand{\ahatd}{\ensuremath{\hat{a}^{\dagger}}\xspace}
\newcommand{\bhat}{\ensuremath{\hat{b}}\xspace}
\newcommand{\bhatd}{\ensuremath{\hat{b}^{\dagger}}\xspace}
\newcommand{\boldr}{\ensuremath{\mathbf{r}}\xspace}
\newcommand{\dr}{\ensuremath{\,d^3\mathbf{r}}\xspace}
\newcommand{\dk}{\ensuremath{\,d^3\mathbf{k}}\xspace}
\newcommand{\etal}{\emph{et al.\/}\xspace}
\newcommand{\ie}{i.e.}
\newcommand{\eq}[1]{Eq.~(\ref{#1})\xspace}
\newcommand{\fig}[1]{Fig.~\ref{#1}\xspace}
\newcommand{\abs}[1]{\left| #1 \right|}
\newcommand{\proj}[2]{\left| #1 \rangle\langle #2\right| \xspace}
\newcommand{\Qhat}{\ensuremath{\hat{Q}}\xspace}
\newcommand{\Qhatd}{\ensuremath{\hat{Q}^\dag}\xspace}
\newcommand{\phihatd}{\ensuremath{\hat{\phi}^{\dagger}}\xspace}
\newcommand{\phihat}{\ensuremath{\hat{\phi}}\xspace}
\newcommand{\boldk}{\ensuremath{\mathbf{k}}\xspace}
\newcommand{\boldp}{\ensuremath{\mathbf{p}}\xspace}
\newcommand{\boldsigma}{\ensuremath{\boldsymbol\sigma}\xspace}
\newcommand{\boldalpha}{\ensuremath{\boldsymbol\alpha}\xspace}
\newcommand{\grad}{\ensuremath{\boldsymbol\nabla}\xspace}
\newcommand{\parti}[2]{\frac{ \partial #1}{\partial #2} \xspace}
 \newcommand{\vs}[1]{\ensuremath{\boldsymbol{#1}}\xspace}
\renewcommand{\v}[1]{\ensuremath{\mathbf{#1}}\xspace}
\newcommand{\Psihat}{\ensuremath{\hat{\Psi}}\xspace}
\newcommand{\Psihatd}{\ensuremath{\hat{\Psi}^{\dagger}}\xspace}
\newcommand{\Vhatd}{\ensuremath{\hat{V}^{\dagger}}\xspace}
\newcommand{\Xhat}{\ensuremath{\hat{X}}\xspace}
\newcommand{\Xhatd}{\ensuremath{\hat{X}^{\dag}}\xspace}
\newcommand{\Yhat}{\ensuremath{\hat{Y}}\xspace}
\newcommand{\Yhatd}{\ensuremath{\hat{Y}^{\dag}}
\xspace}
\newcommand{\jhat}{\ensuremath{\hat{J}}
\xspace}
\newcommand{\lhat}{\ensuremath{\hat{L}}
\xspace}
\newcommand{\Nhat}{\ensuremath{\hat{N}}
\xspace}
\newcommand{\ddt}{\ensuremath{\frac{d}{dt}}
\xspace}
\newcommand{\nset}{\ensuremath{n_1, n_2,\dots, n_k}
\xspace}
\newcommand{\ml}[1]{{\color{blue}#1}}
\newcommand{\cmc}[1]{{\color{red}#1}}
\newcommand{\sah}[1]{{\color{magenta}#1}}

\title{Heisenberg-limited metrology with information recycling}

\author{Simon A.~Haine}
\affiliation{School of Mathematics and Physics,  University of Queensland, Brisbane, QLD, 4072, Australia}
\author{Stuart S.~Szigeti}
\affiliation{School of Mathematics and Physics,  University of Queensland, Brisbane, QLD, 4072, Australia}
\affiliation{ARC Centre for Engineered Quantum Systems, University of Queensland, Brisbane, QLD 4072, Australia}
\author{Matthias D.~Lang}
\affiliation{Center for Quantum Information and Control, University of New Mexico, Albuquerque, New Mexico, USA}
\author{Carlton M.~Caves}
\affiliation{Center for Quantum Information and Control, University of New Mexico, Albuquerque, New Mexico, USA}
\affiliation{ARC Centre for Engineered Quantum Systems, University of Queensland, Brisbane, QLD 4072, Australia}

\begin{abstract}
Information recycling has been shown to improve the sensitivity of atom interferometers by exploiting atom-light entanglement. In this Rapid Communication, we apply information recycling to an interferometer where the input quantum state has been partially transferred from some donor system. We demonstrate that when the quantum state of this donor system is from a particular class of number-correlated Heisenberg-limited states, information recycling yields a Heisenberg-limited phase measurement.  Crucially, this result holds \emph{irrespective\/} of the fraction of the quantum state transferred to the interferometer input and also for a general class of number-conserving quantum-state-transfer processes, including ones that destroy the first-order phase coherence between the branches of the interferometer.  This result could have significant applications in Heisenberg-limited atom interferometry, where the quantum state is transferred from a Heisenberg-limited photon source, and in optical interferometry where the loss can be monitored.
\end{abstract}

\maketitle

When performing an interferometric measurement with a limited number of particles, $N$, there can be significant benefit to using a nonclassical input state to improve the phase sensitivity beyond the standard quantum noise limit (QNL) (shot-noise limit) of $\Delta \phi \sim 1/\sqrt{N}$~\cite{Caves:1981,Bachor2004a}. The ultimate limit to sensitivity is the Heisenberg limit $\Delta \phi \sim 1/N$~\cite{Holland1993a,Giovannetti2006a}. In particular, a Mach-Zehnder (MZ) interferometer can achieve Heisenberg-limited phase sensitivity if the input state has perfect number correlations between the two interferometer modes~\cite{Lucke:2011, Demkowicz-Dobrzanski:2014}.  An example is the two-mode squeezed vacuum state~\cite{Loudon:1987}, which is routinely generated in quantum optics laboratories~\cite{Bachor2004a}.

There exist metrological devices, however, where Heisenberg-limited input states are difficult to generate, such as inertial sensors based on atom interferometry. In such cases, Heisenberg-limited interferometry might still be possible provided a Heisenberg-limited state from a \emph{donor system\/} (e.g., two-mode squeezed optical vacuum) can be mapped to this \emph{acceptor system}. This possibility was demonstrated theoretically in~\cite{Szigeti:2014b}, where quantum state transfer (QST) between squeezed light and atoms was shown to enhance the sensitivity of atom interferometry well below the {QNL}. Similar results are also possible in other contexts, as proposals exist for achieving QST between donor photons and a range of acceptor systems, including atomic motional states~\cite{Parkins:1999}, room-temperature and laser cooled atomic vapours~\cite{Hammerer:2010}, Bose-Einstein condensates of dilute atomic vapors~\cite{Jing:2000, Fleischhauer:2002b, Haine:2005, Haine:2005b, Haine:2006b}, ions~\cite{Stute:2013}, solid state systems~\cite{Hammerer:2010}, and mechanical oscillators~\cite{Zhang:2003}.

Unfortunately, in practice any QST process is imperfect, and even a small degree of imperfection results in a large degradation of the acceptor system's phase sensitivity from the Heisenberg limit~\cite{Demkowicz-Dobrzanski:2012, Szigeti:2014b}. It was first shown in \cite{Haine:2013} that atom-light entanglement can be used to enhance the sensitivity of atom interferometry by applying the technique of information recycling. Furthermore, \cite{Szigeti:2014b, Tonekaboni:2015} revealed that if this atom-light entanglement takes the form of a QST process, then in very specific situations, information recycling can help atom interferometers achieve Heisenberg-limited sensitivities. Here we explicitly prove a generalized version of this result and identify the precise (but still very general) conditions under which it holds. That is, we show that if the donor source displays perfect number correlations, then the acceptor particles give Heisenberg-limited sensitivity \emph{regardless of the QST efficiency} when used in a MZ interferometer, provided information recycling is applied.  This is true regardless of the physical mechanism for QST, provided that the QST process is number conserving.

\emph{Number-correlated MZ interferometer.}  To determine the best phase sensitivity possible for a given interferometry scheme, we appeal to the quantum Fisher information. As discussed in~\cite{Demkowicz-Dobrzanski:2014, Toth:2014}, the quantum Fisher information $\mathcal{F}$ places an absolute lower bound on the phase sensitivity, $\Delta \phi \geq 1/\sqrt{\mathcal{F}}$, called the quantum Cram{\'e}r-Rao bound (QCRB), which applies regardless of the choice of measurement and phase estimation procedure; the bound depends only on the input state.

It is known ~\cite{Demkowicz-Dobrzanski:2014, Toth:2014} that when a pure state is used as the input to a lossless MZ interferometer (\ie, beamsplitter-mirror-beamsplitter configuration), the quantum Fisher information for estimating a differential phase shift is given by $\mathcal{F} = 4 (\langle \hat{L}_y^2\rangle - \langle \hat{L}_y\rangle^2 )$, where $\hat{L}_k \equiv \frac12\v{b}^\dag \sigma_k \v{b}$ defines pseudo-spin operators, $\v{b} = (\bhat_1, \bhat_2)^T$, $\bhat_j$ are the usual bosonic annihilation operators for the two modes, and  $\sigma_k$ are the set of Pauli spin matrices.

Consider now a two-mode state that displays perfect number correlations between the two input modes,
\begin{align}
|\Psi_b\rangle = \sum_{N=0}^\infty c_N |N, N\rangle\,.
\label{eq:Psib}
\end{align}
When used as the input to a MZ interferometer, the quantum Fisher information is given by
\begin{align}
\mathcal{F}_b=\frac{V(\hat N_t)+N_t(N_t+2)}{2}\,,
\label{donor_Fisher}
\end{align}
where $\hat N_t = \bhatd_1\bhat_1 + \bhatd_2\bhat_2$ is the operator for the total number of particles, $N_t=\langle\hat N_t\rangle$ is its expectation value, and $V(\hat X)$ denotes the variance of $\hat X$.  For the twin-Fock state $|\Psi_\text{TF}\rangle = |N/2, N/2\rangle$, the variance is zero, so $\mathcal{F}_b=N_t(N_t+2)/2$.  Two-mode squeezed vacuum \cite{Caves1985a,Schumaker1985a,Schumaker1986a},
\begin{equation}
|\Psi_\text{sq}(r) \rangle
=\text{sech}\,|r| \sum_{N=0}^\infty (-e^{-i\theta}\tanh|r|)^N | N, N \rangle\,,
\end{equation}
with $r = |r|e^{i\theta}$, has variance $V(\hat N_t)=N_t(N_t+2)$ and thus $\mathcal{F}_b=N_t(N_t+2)$.

For a particular choice of measurement signal, $\hat{\mathcal{S}}$, the phase uncertainty is given by $\Delta\phi=\sqrt{\smash[b]{V(\hat{\mathcal{S}})}}/|\partial_\phi\langle\hat{\mathcal{S}}\rangle|$.  Input states of the form~(\ref{eq:Psib}) have no mean field, so the resulting interferometer runs on what would conventionally be called noise; more precisely, they rely on second-order coherence~\cite{Loudon2000a} between the branches of the MZ interferometer, in contrast to the first-order coherence that is required for conventional interferometry.   The signal choice $\hat{\mathcal{S}} = \hat{L}_z^2$ is optimal at the operating points $\phi=0,\pi$, giving a phase uncertainty~\cite{app}
\begin{equation}\label{eq:DeltaphiMZ}
\Delta\phi=\sqrt{\frac{2}{V(\hat N_t)+N_t(N_t+2)}}
\end{equation}
for sensing small changes away from the operating point.  This signal choice thus achieves the {QCRB}.

Since the MZ interferometer does not require first-order coherence between the branches, the phase uncertainty~(\ref{eq:DeltaphiMZ}) is achieved by any input (mixed) state of the form~\cite{app}
\begin{align}
\hat\rho_b=\sum_{M,N=0}^\infty\rho_{MN}|M,M\rangle\langle N,N|\,,
\label{eq:rhob}
\end{align}
not just by the pure states~(\ref{eq:Psib}), for which $\rho_{MN}=c_Mc_N^*$.  We define $p_N \equiv \rho_{NN}$.  When  $\rho_{MN}$ is diagonal, i.e., $\rho_{MN}=p_N\delta_{MN}$, the number correlations between the input branches are purely classical.

\begin{figure}
\includegraphics[width=0.6\columnwidth]{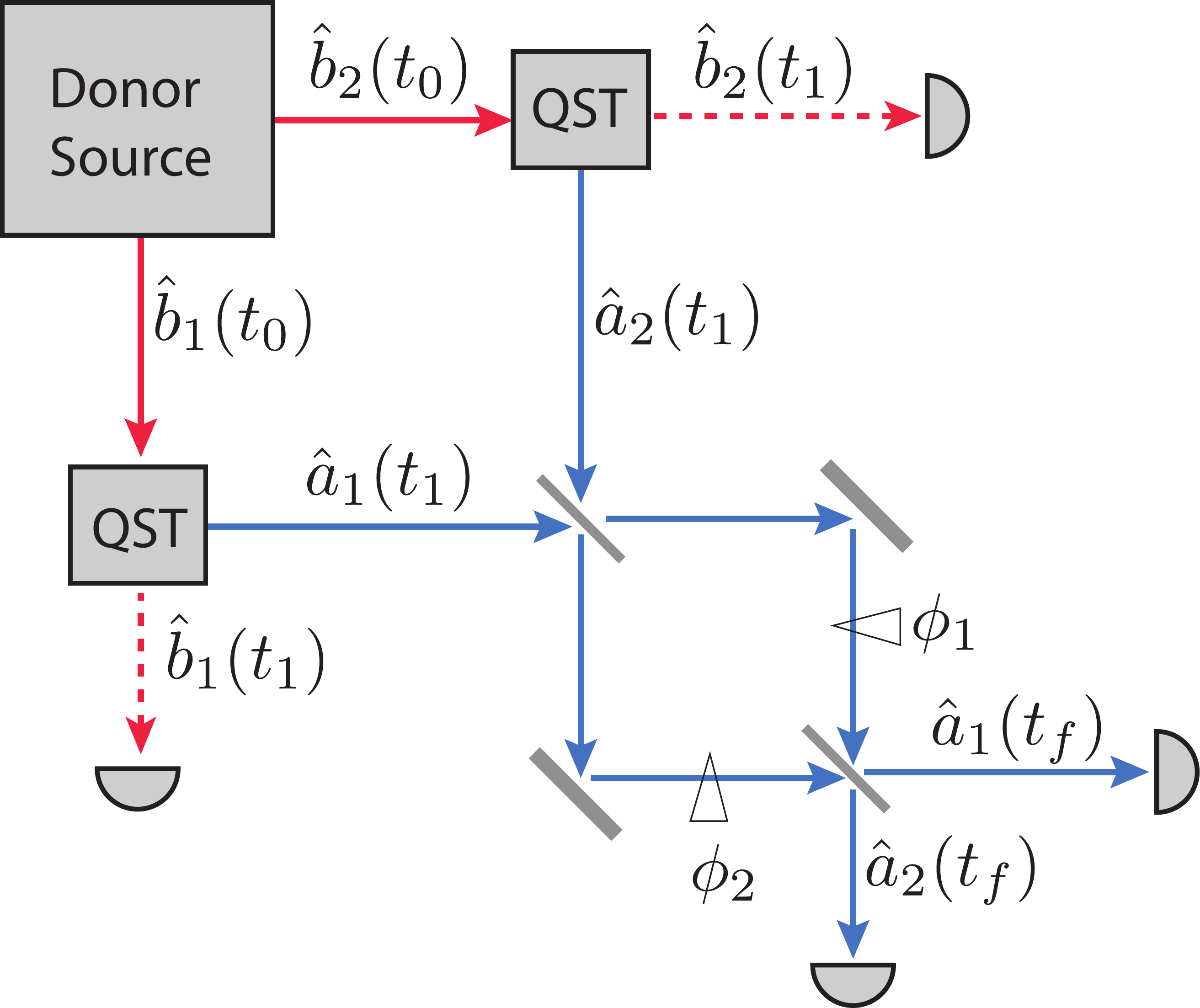}
\caption{Schematic of a donor-enhanced MZ interferometer. Initially, the two-mode donor system (annihilation operators $\bhat_1$ and $\bhat_2$) is prepared in the state $\hat\rho_b$; both modes of the acceptor system (annihilation operators $\ahat_1$ and $\ahat_2$) are initially in vacuum, so we do not depict their inputs to the QST processes in the diagram. Each mode of the donor system undergoes some QST process, transferring part or all of its quantum state to the corresponding mode of the acceptor system at time $t_1$. The two modes of the acceptor system then form the inputs to a conventional MZ interferometer, which is sensitive to the differential phase shift $\phi = \phi_1-\phi_2$. Information recycling is implemented by detecting the number of particles in all four output modes.
}
\label{scheme}
\end{figure}

\emph{Donor-enhanced MZ interferometer.}  Now suppose we want to map the Heisenberg-limited state $\hat\rho_b$ from this donor system to some two-mode acceptor system. This scenario is depicted in Fig.~\ref{scheme}. At $t=t_0$, the quantum state of the system is prepared such that the state of the donor system is $\hat\rho_b$, while the two modes of the acceptor system (annihilation operators $\ahat_1$ and $\ahat_2$) are unoccupied, giving a total state
\begin{equation}
\hat\rho(t_0)\label{eq:rhot0}
=\sum_{M,N=0}^\infty\rho_{MN}|M,0,M,0\rangle\langle N,0,N,0|\,.
\end{equation}
A QST process is implemented such that at $t=t_1$, some or all of the particles are transferred from mode 1(2) of our donor system to mode 1(2) of our acceptor system. The acceptor particles are then used as the input to a MZ interferometer.

A perfect QST process performs the map $|N,0\rangle\rightarrow|0,N\rangle$ in each branch of the interferometer, and consequently the MZ interferometer composed of the two acceptor modes is Heisenberg limited. In practice, however, the QST process is imperfect.  Some particles remain in the donor modes at time $t_1$, and this results in a loss of correlations when considering only the acceptor modes.  As was shown in~\cite{Demkowicz-Dobrzanski:2014, Szigeti:2014b}, even a small loss of correlations can severely degrade sensitivity.  Fortunately, we can reduce this degradation by monitoring those donor particles still remaining after the QST process and incorporating this information as part of our phase-estimation procedure. This technique of \emph{information recycling\/} has been shown to enhance the sensitivity within specific atom interferometric schemes reliant on two-photon Raman transitions~\cite{Haine:2013, Szigeti:2014b}. The surprising result we show here is that a Heisenberg-limited donor source coupled with information recycling yields Heisenberg-limited interferometry with the acceptor modes \emph{irrespective of the QST efficiency\/} or the physical mechanism of the QST process.

To show this, we now consider the state after incomplete QST.  Without specifying the physical mechanism of the QST process, we apply the following physically motivated constraints:
\begin{enumerate}
	\item The QST process occurs in two \emph{independent branches\/}; \ie, donor mode $\hat{b}_1$($\hat{b}_2$) can only exchange particles with acceptor mode $\hat{a}_1$($\hat{a}_2$), and neither branch is affected by the other.
	\item Each branch of the QST process \emph{conserves particle number\/}; \ie, $\bhatd_j\bhat_j + \ahatd_j\ahat_j$ is a conserved quantity for $j= 1,2$.
	\item The QST process is \emph{symmetric\/} with respect to the exchange $\hat{b}_1 \leftrightarrow \hat{b}_2$ and $\hat{a}_1 \leftrightarrow \hat{a}_2$; \ie, the two independent branches of the QST process are identical. \label{sym_cond}
\end{enumerate}
Although a \emph{beamsplitter\/} transformation, and (in the low depletion regime), the atom-light QST process from \cite{Szigeti:2014b}  satisfy these requirements, these conditions are also satisfied by a broad class of QST processes, both unitary and nonunitary. For example, they allow very complicated QST processes where the QST Hamiltonian contains higher-order couplings; heuristically, this might result in a QST efficiency that depends on the number of particles in the donor mode.  Furthermore, the constraints allow for situations where the QST process is mediated by some other set of modes $\hat{c}_k$ (e.g., a reservoir), which might be depleted and thus reduce the QST efficiency as more particles are transferred, as seen in~\cite{Szigeti:2014b}. A somewhat fanciful, but certainly not the most general Hamiltonian that satisfies the constraints of such a QST process is
\begin{align}
\hat{H}\!=\!
\sum_{i=1,2}
\sum_{\substack{n, m, l \\ q, p, k}}\!h^{nml}_{qpk}\!
\left[ (\hat{a}_i^\dag \hat{a}_i)^n (\hat{b}_i^\dag \hat{b}_i)^m (\hat{a}_i^\dag \hat{b}_i)^l (\hat{c}_{i,k}^\dag)^q \hat{c}_{i,k}^p + h.c.\right]\,.
\end{align}

A general QST process that satisfies the conditions 1--3 performs the following map in each branch:
\begin{equation}\label{eq:Amap}
|M,0\rangle\langle N,0|\rightarrow
\sum_{m=0}^M\sum_{n=0}^N
A_{Mm,Nn}|M-m,m\rangle\langle N-n,n|\,.
\end{equation}
There are no constraints on $A_{Mm, Nn}$ other than the usual physical constraints of normalization and complete positivity.  $P_{n|N} \equiv A_{Nn,Nn}$ is the conditional probability that there are $n$ particles in an acceptor mode, given $N$ particles initially in the corresponding donor mode. 

Under the QST map~(\ref{eq:Amap}), the state $\hat\rho(t_0)$ of Eq.~(\ref{eq:rhot0}) is mapped to the (generally mixed) state
\begin{align}
&\hat\rho(t_1)=
\sum_{M,N=0}^\infty\rho_{MN}\sum_{\substack{m_1,n_1\\m_2,n_2}}
A_{Mm_1,Nn_1}A_{Mm_2,Nn_2}\nonumber\\
&\times|M-m_1,m_1,M-m_2,m_2\rangle\langle N-n_1,n_1,N-n_2,n_2|\,.
\label{rhodef}
\end{align}
Notice that we only require that number correlations between the branches be maintained; dephasing within or between the branches is perfectly acceptable.

Introducing the pseudo-spin operators for the acceptor modes, $\jhat_k \equiv \frac12\v{a}^\dag \sigma_k \v{a}$, where $\v{a} = (\ahat_1, \ahat_2)^T$, the unitary operator for the MZ interferometer performs the following transformations: $\hat{J}_z(t_f) = \hat{U}^\dag \jhat_z(t_1)\hat{U} = \hat{J}_z(t_1)\cos\phi - \hat{J}_x(t_1)\sin\phi$, and  $\hat{L}_z(t_f) =\hat{U}^\dag \hat{L}_z(t_1)\hat{U} =  \hat{L}_z(t_1)$, since only the acceptor particles take part in the interferometric process. As in~\cite{Szigeti:2014b}, we estimate the phase by measuring the number of particles at the four output ports (see Fig.~\ref{scheme}) and constructing the signal $\hat{\mathcal{S}} = [\jhat_z(t_f) + \hat{L}_z(t_f)]^2$.  Although only $\jhat_z$ contains phase information, the noise in $\jhat_z$ is anticorrelated with $\lhat_z$, so measuring both quantities allows us to correct for this noise and therefore improve sensitivity.

To evaluate the phase sensitivity, we need the first and second moments of $\hat{\mathcal{S}}$ in the state~(\ref{rhodef}).  Since the QST process and the angular-momentum operators preserve total particle number, there is no interference between sectors with different numbers of particles; the desired moments are averages over $p_N=\rho_{NN}$.  The anticorrelation of $\jhat_z$ and $\lhat_z$, expressed by $\jhat_z\hat\rho(t_1)=-\lhat_z\hat\rho (t_1)$, allows us to convert $\lhat_z$ in these moments to $\jhat_z$.  The anticorrelation implies that $\hat\rho(t_1)$ is invariant under rotations about the $z$ axis; in particular, a rotation by $\pi$, which takes $\jhat_x$ to $-\jhat_x$, implies that all terms with an odd number of $\jhat_x$ operators have vanishing expectation value.  At the most sensitive operating point, $\phi=0$, the phase sensitivity is~\cite{app}
\begin{equation}
\Delta\phi
=\frac{\sqrt{V(\hat{\mathcal{S}})}}{\abs{\partial_\phi\langle\hat{\mathcal{S}}\rangle}}
= \frac{1}{2\langle\jhat_x^2\rangle^{1/2}}
= \sqrt{\frac{1}{2\langle \Nhat_1\Nhat_2\rangle + N_a}} \, ,
\label{eq:DeltaphiIR}
\end{equation}
where $\Nhat_j = \ahatd_j(t_1)\ahat_j(t_1)$, and $N_a = \langle \Nhat_1 + \Nhat_2\rangle$ is the average number of acceptor particles and thus the number of particles that take part in the interferometric process.

We can put a lower bound on $\langle \Nhat_1\Nhat_2\rangle$ by noting that a state of the form~(\ref{rhodef}) gives
\begin{align}
\langle\Nhat_1\Nhat_2\rangle
=\sum_{N=0}^\infty p_N\langle \Nhat_1\rangle_N \langle \Nhat_2\rangle_N
=\sum_{N=0}^\infty p_N\langle \Nhat_1\rangle_N^2\,.
\end{align}
Here $\langle\hat{N}_j\rangle_N=\sum_{n_j=0}^N n_j P_{n_j|N}$ is the \emph{conditional expectation value\/} of the number of particles in acceptor mode~$j$, given $N$ initial particles in donor mode~$j$. That the conditional probabilities are the same in the two branches ensures that $\langle \Nhat_1\rangle_N = \langle \Nhat_2\rangle_N$.  Convexity implies that
\begin{equation}
\langle\Nhat_1\Nhat_2\rangle\ge
\left(\sum_{N=0}^\infty p_N \langle \hat{N}_1\rangle_N\right)^2
=\langle \hat{N}_1\rangle^2=\frac{1}{4}N_a^2\,,
\end{equation}
which gives an upper bound on the phase sensitivity of any QST process applied to the initial state~$\hat{\rho}(t_0)$,
\begin{equation}
\Delta \phi\leq\sqrt{\frac{2}{N_a(N_a + 2)}} \simeq \frac{\sqrt{2}}{N_a}\,.
\label{phi_bound}
\end{equation}

The important feature of this result is that the Heisenberg limit is recovered, with respect to the number of particles, $N_a$, taking part in the interferometer, rather than the total number of particles $N_t$. Although the absolute sensitivity is less than with perfect QST, this is purely due to loss of particles, rather than to loss of correlations.  We stress that this is not the \emph{true\/} Heisenberg limit, in the sense that we have used $N_t \geq N_a$ particles to make the measurement, but only $N_a$ of them have passed through the interferometer.  Without the application of information recycling, however, the sensitivity is significantly worse than $1/N_a$~\cite{app}.

For the specific case when the donor source is a twin-Fock state, $|\Psi_b\rangle=|\Psi_\text{TF}\rangle$, we get $\langle \Nhat_1 \Nhat_2\rangle = \langle \Nhat_1 \rangle\langle \Nhat_2\rangle$, which gives a phase sensitivity that saturates the bound (\ref{phi_bound}) and is entirely independent of the QST efficiency or even the form of the number-conserving QST interaction.  For other initial states, there might be a weak dependence on the QST process (as seen for the beamsplitting case below); nevertheless the phase sensitivity is guaranteed to be at least as good as that given by the twin-Fock state.  To be more quantitative about the performance of states other than $|\Psi_\text{TF}\rangle$, we need to specify a particular Hamiltonian governing the QST process.

\emph{Beamsplitter QST process.}  We now consider the simplest possible QST process, a beamsplitter.  The Hamiltonian describing this process, $\hat{H} \propto \sum_{j=1,2}(\ahat_j\bhatd_j + \ahatd_j\bhat_j)$, leads to the unitary transformation
\begin{subequations}
\label{BS_relations}
\begin{align}
	\hat{a}_j(t_1)	&= \sqrt{1-\mathcal{Q}}\,\hat{a}_j(t_0) - i\sqrt{\mathcal{Q}}\,\hat{b}_j(t_0)\,, \\
	\hat{b}_j(t_1)	&= \sqrt{1-\mathcal{Q}}\,\hat{b}_j(t_0) - i\sqrt{\mathcal{Q}}\,\hat{a}_j(t_0)\,.
\end{align}
\end{subequations}
Here $\mathcal{Q}$ is the \emph{QST efficiency}, \ie, the fraction of donor particles mapped to the acceptor modes.

The transformation~(\ref{BS_relations}) allows us to evaluate \eq{eq:DeltaphiIR} explicitly to determine the precise dependence on the QST efficiency.  With the initial state~(\ref{eq:rhot0}), we get $\langle \hat{N}_1\hat{N}_2\rangle = \big(\mathcal{Q}^2V(\hat N_t)+ \langle N_a\rangle^2\big)/4$, and the phase sensitivity in the presence of information recycling is
\begin{equation}
\Delta \phi = \sqrt{\frac{2}{\mathcal{Q}^2 V(\hat N_t) + N_a(N_a+2)}}\,.
\label{delta_phi_BS}
\end{equation}
For a twin-Fock input, which has $V(\hat N_t)=0$, the phase sensitivity does not depend on $\mathcal{Q}$ and is given by the bound in Eq.~(\ref{phi_bound}).  When the donor state is two-mode squeezed vacuum, $|\Psi_b\rangle = |\Psi_\text{sq}\rangle$, we find that $\Delta \phi = 1/\sqrt{N_a(N_a+1+\mathcal{Q})}$, which has only a weak dependence on $\mathcal{Q}$.  Indeed, it is clear that to leading order in the total number of acceptor particles, $N_a=\mathcal{Q}N_t$, the sensitivity~(\ref{delta_phi_BS}) has Heisenberg scaling for any donor input state~(\ref{eq:rhob}), \emph{regardless of the QST efficiency~$\mathcal{Q}$}.  This gives a clear illustration of the power of information recycling as a tool to enable quantum metrology.

It is instructive to compute the quantum Fisher information~$\mathcal{F}_a$ for the donor-acceptor interferometer.  With the pure initial state~(\ref{eq:Psib}) and a beamsplitter QST process, the state remains pure, and the quantum Fisher information is simply $\mathcal{F}_a = 4[\langle \jhat_y(t_1)^2\rangle - \langle \jhat_y(t_1)\rangle^2]$.  The transformations~(\ref{BS_relations}) allow us to compute these expectations with respect to the initial state. Since the acceptor modes are initially vacuum, we obtain
\begin{equation}
\mathcal{F}_a
= \mathcal{Q}^2 \mathcal{F}_b + \left(1-\mathcal{Q}\right)N_a = \frac{\mathcal{Q}^2 V(\hat N_t) + N_a(N_a+2)}{2}\,.
\label{Eq_Fa}
\end{equation}
Comparing with the sensitivity~(\ref{delta_phi_BS}), it is clear that our information-recycled signal achieves the best possible Heisenberg scaling, i.e., by saturating the~{QCRB}.

\begin{figure}
\includegraphics[width=\columnwidth]{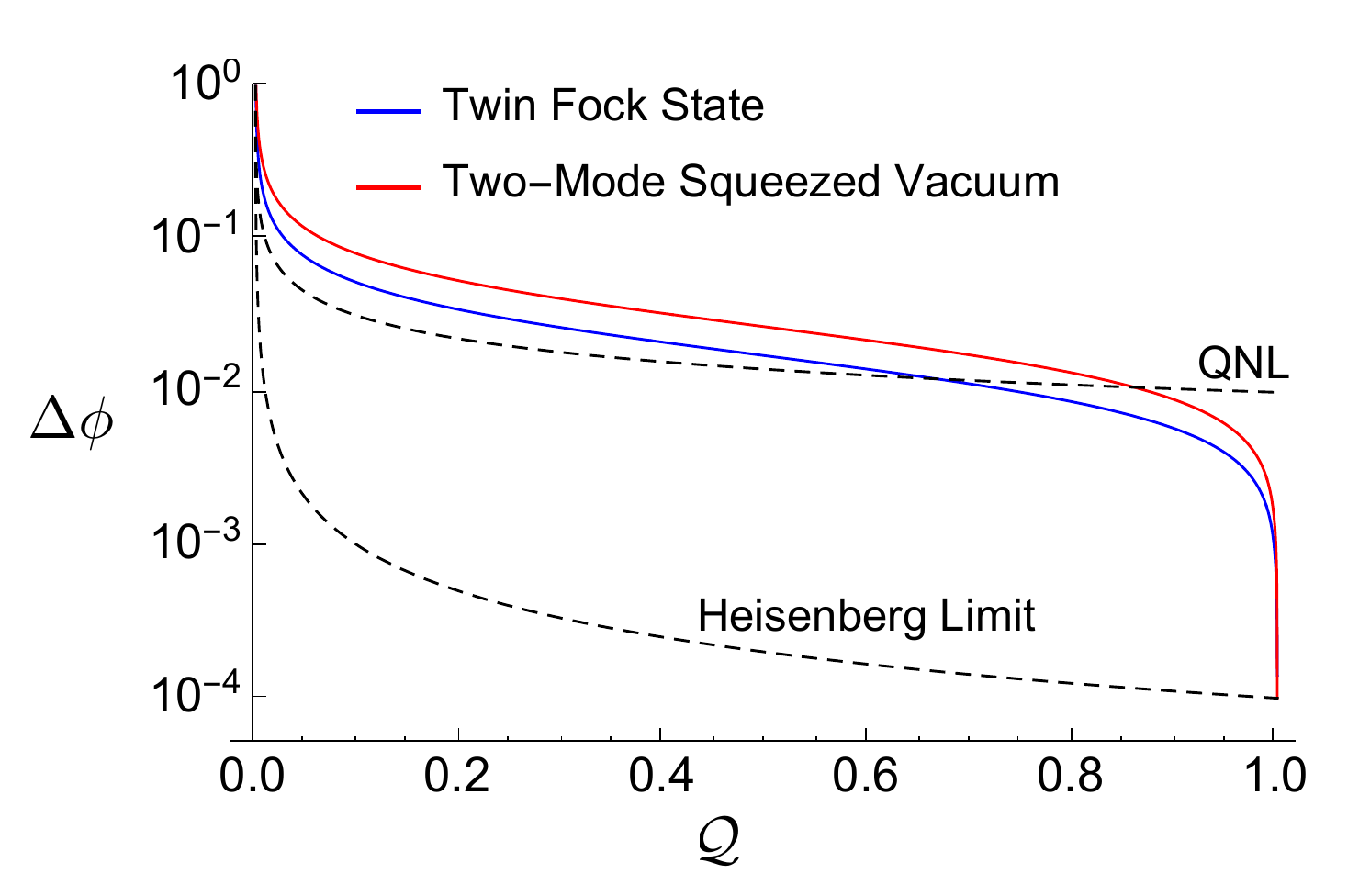}
\caption{Examples of the QST dependence of the phase sensitivity at the optimal operating point without information recycling (i.e., $\hat{\mathcal{S}}=\hat{J}_z^2$), assuming a beamsplitter QST process, for initial donor states $|\Psi_\text{TF}\rangle$ (solid blue) and $|\Psi_\text{sq}\rangle$ (solid red) and using $N_t = 10^4$. The upper and lower dashed lines mark the standard QNL, $\Delta \phi = 1/\sqrt{N_a}$, and the Heisenberg limit, $\Delta \phi = 1/N_a$, respectively. Heisenberg scaling is rapidly lost for small departures from perfect QST.  In contrast, the sensitivity~(\ref{delta_phi_BS}) with information recycling has Heisenberg scaling $\propto1/N_a$ for all $\mathcal{Q}$. The analytic expressions for the sensitivity $\Delta\phi$, as a function of $\phi$ and at the optimal operating point, are in the Supplemental Material~\cite{app}.}
\label{fig_no_info}
\end{figure}

In contrast to these results, when information recycling is not applied, the beamsplitter QST process acts as a linear loss mechanism and Heisenberg scaling is lost (see Fig.~\ref{fig_no_info}).  This loss of Heisenberg scaling occurs for relatively small deviations of $\mathcal{Q}$ from perfect QST and affects any initial state of the form~(\ref{eq:rhob})~\cite{app} (see also \cite{Lang:2014, escher2011general}).

\begin{figure}
\includegraphics[width=0.8\columnwidth]{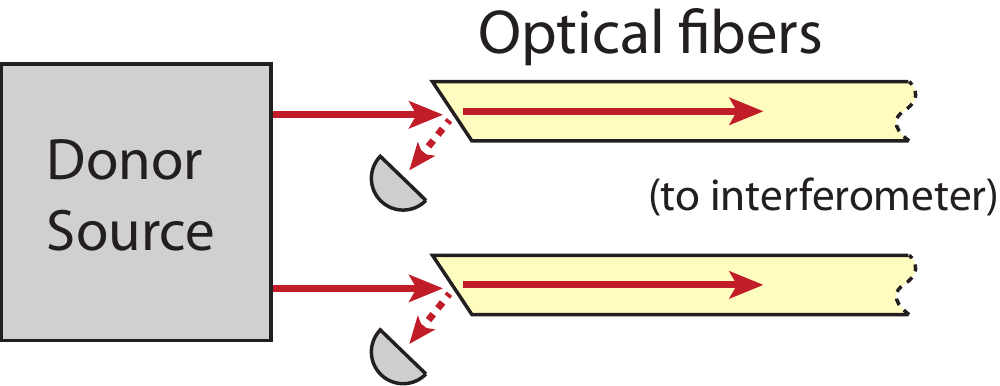}
\caption{Free-space photons from a donor source are coupled into optical fibers, which form the arms of a MZ interferometer. By measuring scattered photons, information recycling can be used to counteract the deleterious effects of inefficient coupling between the donor photon source and the optical fibers.}
\label{fig_fiber}
\end{figure}

\emph{Applications.}   Donor-enhanced interferometry with information recycling requires the following: (i)~a correlated source of donor particles, (ii)~partial QST between the donor particles and some acceptor system that operates in two independent and symmetric branches,  and (iii)~the ability to detect both donor and acceptor particles. It might be particularly useful in situations where there are abundant donor particles and a limited number of acceptor particles [such as QST from photons (donor) to atoms (acceptor) for the purposes of atom interferometry], since the QST efficiency becomes irrelevant once $N_a$ equals the total number of available acceptor particles. In addition to Heisenberg-limited atom interferometry, another potential application for this scheme is optical interferometry which requires coupling into optical fibers before an interferometer (Fig.~\ref{fig_fiber}). Here, coupling between the freely propagating modes (donor system) and the fiber modes (acceptor system) represents the QST process. Typically there will be some scattering into other modes, which is a source of inefficient QST. Information recycling could be implemented by detecting the scattered photons. Since our scheme only requires photon counting, rather than homodyne detection, information recycling could still be implemented even if the scattering is incoherent, and into a range of spatial modes. 

\emph{Acknowledgements.} We acknowledge useful discussions with W.~Bowen, J.~Combes, J.~Corney, M.~Davis, and T.~Stace.  This work was supported in part by ARC Project No.~DE130100575 and by NSF Grant No.~PHY-1314763 and No.~PHY-1212445.  SSS acknowledges the support of I.~P. McCulloch and the ARC Centre of Excellence for Engineered Quantum Systems (Project No.~CE110001013).

\bibliography{info_recyc_HB_bib}

\begin{widetext}

\section{Supplementary Material}

\subsection{Phase sensitivity of Mach-Zehnder interferometer with number-correlated input}

For a MZ interferometer with number-correlated input
\begin{equation}
\hat\rho_b=\sum_{M,N}\rho_{MN}|M,M\rangle\langle N,N|\,,
\end{equation}
an optimal signal is $\hat{\mathcal{S}}=\hat L_z^2$.  The uncertainty in the measured phase is
\begin{equation}\label{eq:Deltaphi}
\Delta\phi=\frac{\sqrt{V(\hat{\mathcal{S}})}}{|\partial_\phi\langle\hat{\mathcal{S}}\rangle|}\,.
\end{equation}
The interferometer transforms $\hat L_z$ to $\hat L_z\cos\phi-\hat L_x\sin\phi$, so the moments of the signal are given by expectation values of even powers of $\hat L_z\cos\phi-\hat L_x\sin\phi$ in the state $\hat\rho_b$.  Because $\hat L_z$ and $\hat L_x$ preserve the total photon number, only the diagonal terms of the density operator contribute to the relevant moments, which are moments of $N$ with respect to $p_N=\rho_{NN}$.  In addition, we have $\hat L_z\hat\rho_b=0=\hat\rho_b\hat L_z$, which implies that $\hat\rho_b$ is invariant under rotations about the $z$ axis; in particular, a rotation by $\pi$, which takes $\hat L_x$ to $-\hat L_x$ shows that moments that have an odd power of $\hat L_x$ vanish.

Using these facts to calculate the expectation value and variance of $\hat{\mathcal{S}}$ gives
\begin{align}
\langle\hat{\mathcal{S}}\rangle&=\langle  \hat L_x^2\rangle\sin^2\!\!\phi\,,\\
V(\hat{\mathcal{S}})
&=\langle  \hat L_x\hat L_z^2\hat L_x\rangle\sin^2\!\!\phi\cos^2\!\!\phi+V(\hat L_x^2)\sin^4\!\!\phi\,,
\end{align}
which leads to a measured-phase variance
\begin{equation}
(\Delta\phi)^2=\frac14\left(
\frac{\langle  \hat L_x\hat L_z^2\hat L_x\rangle}{\langle  \hat L_x^2\rangle^2}+
\frac{V(\hat L_x^2)}{\langle  \hat L_x^2\rangle^2}\tan^2\!\!\phi
\right)\,.
\end{equation}
The optimal operating points are $\phi=0,\pi$.

The angular-momentum moments are
\begin{subequations}
\begin{align}
\langle  \hat L_x^2\rangle&=\frac12\sum_N N(N+1)p_N=
\frac{\langle\hat N_t^2\rangle+2\langle\hat N_t\rangle}{8}\,,\\
\langle\hat L_x\hat L_z^2\hat L_x\rangle&=
\frac12\sum_N N(N+1)p_N=\langle\hat L_x^2\rangle\,,\\
V(\hat L_x^2)&=\sum_N\bigg(\frac38 N^4+\frac34 N^3+\frac18N^2-\frac14N\bigg)p_N
-\langle\hat L_x^2\rangle^2\nonumber\\
&=\frac{3}{128}\langle\hat N_t^2\rangle+\frac{3}{32}\langle\hat N_t^3\rangle
+\frac{1}{32}\langle\hat N _t^2\rangle - \frac{1}{8}\langle\hat N_t\rangle-\langle\hat L_x^2\rangle^2\,.
\end{align}
\end{subequations}
Therefore, at the optimal operating points the phase sensitivity is
\begin{equation}
\Delta\phi=\frac{1}{2\langle\hat L_x^2\rangle^{1/2}}
=\sqrt{\frac{2}{\langle\hat{N}_t^2\rangle+2\langle\hat N_t\rangle}}\,.
\end{equation}

\subsection{Phase sensitivity for the QST process in the absence of information recycling}

The donor-acceptor state after the QST process (i.e. at time $t_1$) is
\begin{align}
&\hat\rho(t_1)=
\sum_{M,N=0}^\infty\rho_{MN}\sum_{\substack{m_1,n_1\\m_2,n_2}}
A_{Mm_1,Nn_1}A_{Mm_2,Nn_2}
|M-m_1,m_1,M-m_2,m_2\rangle\langle N-n_1,n_1,N-n_2,n_2|\,.
\label{rhodef}
\end{align}
Without information recycling, the signal we measure is $\hat{\mathcal{S}}=[\jhat_z(t_f)]^2$, where
\begin{align}
\jhat_z(t_f) = U_\text{MZ}^{\dagger} \jhat_z(t_1) U_\text{MZ} = \jhat_z(t_1) \cos \phi - \jhat_x(t_1) \sin \phi\,.
\end{align}
In what follows, we condense the notation by writing $\jhat_z \equiv \jhat_z(t_1)$ and $\jhat_x \equiv \jhat_x(t_1)$.

In order to calculate the phase sensitivity $\Delta\phi$ of Eq.~(\ref{eq:Deltaphi}), we need to evaluate the first and second moments of $\hat{\mathcal{S}}$.  Since the QST process that leads to the state~(\ref{rhodef}) and the angular-momentum operators preserve total particle number, there is no interference between sectors with different numbers of particles; the desired moments are thus averages over $p_N=\rho_{NN}$.  Moreover, the anticorrelation of $\jhat_z$ and $\lhat_z$, expressed by $(\jhat_z+\lhat_z)\hat\rho(t_1)=0=-\hat\rho(t_1)(\jhat_z+\lhat_z)$, means that $\hat\rho(t_1)$ is invariant under rotations about the $z$ axis; in particular, a rotation by $\pi$, which takes $\jhat_x$ to $-\jhat_x$, implies that all terms with an odd number of $\jhat_x$ operators have vanishing expectation value.

Generally, we have
\begin{align}
\hat{\mathcal{S}}=
[\jhat_z(t_f)]^2
&=\jhat_z^2 \cos^2\! \phi + \jhat_x^2 \sin^2\! \phi - \cos \phi \sin \phi (\jhat_z \jhat_x+ \jhat_x \jhat_z)\,,\\
\hat{\mathcal{S}}^2=[\jhat_z(t_f)]^4
&=\jhat_z^4 \cos^4\!\phi + \jhat_z^2 \jhat_x^2 \cos^2\!\phi \sin^2\!\phi - \jhat_z^2(\jhat_z \jhat_x+ \jhat_x \jhat_z) \cos^3\phi \sin\phi \notag\\
  &\quad+\jhat_x^2 \jhat_z^2 \cos^2\!\phi \sin^2\!\phi + \jhat_x^4 \sin^4\!\phi - \jhat_x^2(\jhat_z \jhat_x+ \jhat_x \jhat_z) \cos\phi \sin^3\!\phi \notag\\
  &\quad-(\jhat_z \jhat_x+ \jhat_x \jhat_z)\jhat_z^2 \cos^3\phi \sin\phi- (\jhat_z \jhat_x+ \jhat_x \jhat_z)\jhat_x^2 \cos\phi \sin^3\!\phi\nonumber\\
  &\quad+ (\jhat_z \jhat_x+ \jhat_x \jhat_z)^2 \cos^2\!\phi \sin^2\!\phi\,.
\end{align}
Applying our rules, we get
\begin{align}
\langle\hat{\mathcal{S}}\rangle
&=\langle\jhat_z^2\rangle\cos^2\! \phi
+ \langle\jhat_x^2\rangle\sin^2\! \phi\,,\\
\langle\hat{\mathcal{S}}^2\rangle
&=\langle  \jhat_z^4\rangle\cos^4\!\phi
+ \langle  \jhat_x^4\rangle\sin^4\!\phi
+\big[\langle\jhat_z^2 \jhat_x^2+\jhat_x^2\jhat_z^2\rangle
+\langle  (\jhat_x\jhat_z + \jhat_z\jhat_x)^2\rangle\big]\cos^2\!\phi \sin^2\!\phi\,,
\end{align}
which gives the squared phase sensitivity
\begin{align}
(\Delta \phi)^2	
= \frac{V(\jhat_z^2)\cot^2\!\phi + V(\jhat_x^2)\tan^2\!\phi
+ C(\jhat_x^2, \jhat_z^2) + \langle   (\jhat_x\jhat_z + \jhat_z\jhat_x)^2 \rangle}
{4\big(\langle  \jhat_z^2 \rangle - \langle  \jhat_x^2 \rangle\big)^2}\,.
\label{eq:sens}
\end{align}
Here $V(\hat X) = \langle   \hat{X}^2 \rangle - \langle   \hat{X} \rangle^2$ is, as throughout, the variance of $\hat X$, and $C(\hat X,\hat Y) = \langle   \hat{X} \hat{Y} + \hat{Y} \hat{X} \rangle - 2\langle   \hat{X} \rangle \langle   \hat{Y} \rangle$ is the symmetrized covariance of $\hat X$ and $\hat Y$.

We can write these expectation values in terms of the number operators for the two acceptor modes, $\hat{N}_1$ and $\hat{N}_2$.  We again make use of the form of the input state and the restrictions on the state transfer, which make any term vanish whose number of creation operators, $\ahatd_1$ or $\ahatd_2$, does not match the corresponding number of annihilation operators, $\ahat_1$ or $\ahat_2$.  The relevant angular-momentum moments are
\begin{subequations}
\begin{align}
\langle   \jhat_z^2\rangle &= \frac{1}{4} \big\langle  (\hat{N}_1 - \hat{N}_2)^2 \big\rangle \,,\\
\langle   \jhat_x^2\rangle &= \langle   \jhat_y^2\rangle
= \frac{1}{4} \big\langle  2 \hat{N}_1 \hat{N}_2 + \hat{N}_1+ \hat{N}_2\big\rangle\,,\\
\langle   \jhat_z^4\rangle &= \frac{1}{16} \big\langle  (\hat{N}_1 - \hat{N}_2)^4 \big\rangle\,,\\
\langle   \jhat_x^4\rangle &= \frac{1}{16} \big\langle  (2 \hat{N}_1 \hat{N}_2 + \hat{N}_1+ \hat{N}_2 + \ahatd_1\ahatd_1\ahat_2 \ahat_2+\ahat_1 \ahat_1 \ahatd_2\ahatd_2)^2\big\rangle\notag\\
&=\frac{1}{16}\big\langle(2 \hat{N}_1 \hat{N}_2 + \hat{N}_1+ \hat{N}_2)^2 + \ahatd_1 \ahatd_1 \ahat_1 \ahat_1
\ahat_2 \ahat_2 \ahatd_2 \ahatd_2+\ahat_1 \ahat_1 \ahatd_1\ahatd_1 \ahatd_2\ahatd_2 \ahat_2 \ahat_2\big\rangle\notag\\
&=\frac{1}{16}\big\langle(2\hat{N}_1 \hat{N}_2 + \hat{N}_1+ \hat{N}_2)^2 + (\hat{N}_1^2-\hat{N}_1)(\hat{N}_2^2+3\hat{N}_2 +2)\notag\\
&\qquad\qquad+(\hat{N}_1^2+3\hat{N}_1 +2)(\hat{N}_2^2-\hat{N}_2)\big\rangle\,,\\
\langle\jhat_z^2 \jhat_x^2\rangle &=  \langle   \jhat_x^2 \jhat_z^2\rangle=\langle   \jhat_z \jhat_x^2 \jhat_z\rangle
=\frac{1}{16}\big\langle(\hat{N}_1-\hat{N}_2)^2(2 \hat{N}_1 \hat{N}_2 + \hat{N}_1+ \hat{N}_2)\big \rangle \,,\\
\langle\jhat_x \jhat_z^2 \jhat_x\rangle\ &= \langle\jhat_z^2 \jhat_x^2\rangle -  \langle   \jhat_z^2\rangle +\langle   \jhat_x^2\rangle\,,\label{Jxzzx}\\
\langle\jhat_x \jhat_z \jhat_x \jhat_z\rangle
&=\langle \jhat_z \jhat_x \jhat_z \jhat_x\rangle\ = \langle\jhat_z^2 \jhat_x^2\rangle - \frac{1}{2}  \langle \jhat_z^2\rangle \label{Jxzxz}\,.
\end{align}
\end{subequations}
As in the main text, we introduce \textit{conditional\/} expectation values to write the moments of the signal~$\hat{\mathcal{S}}$:
\begin{align}
\langle\hat{\mathcal{S}}\rangle
&=\sum_{N=0}^\infty p_N \bigg(\frac{1}{2} \sin^2\!\phi \langle   \hat{N}_1\rangle_N + \frac{1}{2} \cos^2\!\phi \langle   \hat{N}_1^2\rangle_N+\frac{1}{2} (\sin^2\!\phi -\cos^2\!\phi)
\langle\hat{N}_1\rangle_N^2\bigg)\,,\\
\langle   \hat{\mathcal{S}}^2 \rangle
&=\sum_{N=0}^\infty p_N\Bigg(
\left(\frac{1}{2}\cos^2\!\phi\sin^2\!\phi - \frac{\sin^4\!\phi}{4}\right)\langle\hat{N}_1\rangle_N
+\left(-\cos^2\!\phi\sin^2\!\phi + \frac{3\sin^4\!\phi}{8}\right)\langle\hat{N}_1^2\rangle_N\notag\\
&\quad+\frac{3}{4}\cos^2\!\phi\sin^2\!\phi\langle\hat{N}_1^3\rangle_N
+\frac{1}{8} \cos^4\!\phi \langle   \hat{N}_1^4\rangle_N
+\left(\frac{3}{2} \cos^2\!\phi \sin^2\!\phi-\frac{\sin^4\!\phi}{4}\right)\langle\hat{N}_1\rangle_N \langle\hat{N}_1\rangle_N\notag\\
&\quad+\left(-\frac{3}{4} \cos^2\!\phi \sin^2\!\phi + \frac{3 \sin^4\!\phi}{4}\right)
\langle\hat{N}_1^2\rangle_N \langle   \hat{N}_1\rangle_N
+\left(-\frac{1}{2} \cos^4\!\phi+ \frac{3}{2} \cos^2\!\phi \sin^2\!\phi\right)
\langle\hat{N}_1^3\rangle_N\langle   \hat{N}_1\rangle_N \notag\\
&\quad+\left(\frac{3\cos^4\!\phi}{8}-\frac{3}{2} \cos^2\!\phi \sin^2\!\phi+ \frac{3 \sin^4\!\phi}{8}\right)
\langle\hat{N}_1^2\rangle_N\langle\hat{N}_1^2\rangle_N \Bigg)\,.
\end{align}
Here we use the fact that the conditional expectation values are the same in the two branches of the interferometer to convert all the conditional expectation values to mode~1.

Now we specialize to the case where the QST process is a beamsplitter, with $\mathcal{Q}$ denoting the QST efficiency.  In this case the conditional probability distributions in the two branches are binomial distributions, and we can write the number moments in terms of moments $\langle\hat N_t^k\rangle$ of the total particle number in the initial state. As in the main text, we use $N_t \equiv \langle\hat{N}_t\rangle$ to denote the average total particle number.  The relevant moments take the following forms:
\begin{subequations}
\label{J_identities}
\begin{align}
\langle   \jhat_z^2\rangle &= \frac{1}{4} \mathcal{Q} (1-\mathcal{Q}) N_t\,,\\
\langle   \jhat_x^2\rangle &= \langle   \jhat_y^2\rangle = \frac{\mathcal{Q}}{4} \left(\frac{\mathcal{Q} \langle \hat{N}_t^2 \rangle }{2} + N_t \right)\,,\\
\langle   \jhat_z^4\rangle
&= \frac{\mathcal{Q}(1-\mathcal{Q})}{16}\left(3 \mathcal{Q} (1 - \mathcal{Q}) \langle   \hat{N}_t^2 \rangle + \left(6 \mathcal{Q}^2-6 \mathcal{Q}+1\right) N_t \right),\\
\langle   \jhat_x^4\rangle &= \frac{\mathcal{Q}}{16} \left(\frac{3 \mathcal{Q}^3 \langle   \hat{N}_t^4 \rangle }{8} + \frac{3 \mathcal{Q}^2}{2} (2 - \mathcal{Q}) \langle \hat{N}_t^3 \rangle +\frac{\mathcal{Q}}{2} \left(3
   \mathcal{Q}^2-12 \mathcal{Q}+10\right) \langle \hat{N}_t^2 \rangle + \left(1 - 3\mathcal{Q}\right) N_t \right), \\
      \langle   \jhat_z^2 \jhat_x^2\rangle &=  \langle   \jhat_x^2 \jhat_z^2\rangle=\langle   \jhat_z \jhat_x^2 \jhat_z\rangle = \frac{\mathcal{Q}(1-\mathcal{Q})}{16}\left( \frac{\mathcal{Q}^2}{2} \langle \hat{N}_t^3 \rangle + \mathcal{Q} (1 - \mathcal{Q}) \langle \hat{N}_t^2 \rangle + (1-2\mathcal{Q}) N_t \right).
\end{align}
\end{subequations}
This implies that
\begin{align}
\langle\hat{\mathcal{S}}\rangle 
&= \frac{\mathcal{Q}}{4} \left( \frac{\mathcal{Q}}{2} \sin^2\!\phi  \langle   \hat{N}_t^2\rangle + \left[(1-\mathcal{Q}) \cos^2\!\phi + \sin^2\!\phi \right]N_t \right),\\
\langle\hat{\mathcal{S}}^2\rangle &= \frac{3}{128} \mathcal{Q}^4 \sin ^4\!\phi  \langle   \hat{N}_t^4\rangle -\frac{3}{64} \mathcal{Q}^3 \sin ^2\!\phi  (\mathcal{Q} \cos 2 \phi -4 +3 \mathcal{Q})  \langle   \hat{N}_t^3\rangle \notag \\
&\quad+\frac{1}{256} \mathcal{Q} ^2 \left(  39 \mathcal{Q}^2 - 96 \mathcal{Q} + 64 - 4(4 - 3 \mathcal{Q}^2)\cos 2\phi - 3 \mathcal{Q}^2 \cos 4\phi \right) \langle \hat{N}_t^2\rangle \notag \\
&\quad-\frac{1}{16} \mathcal{Q}  \left(\mathcal{Q}  \left[6 \mathcal{Q} \cos^2\!\phi \left(\mathcal{Q}  \cos ^2\!\phi -2\right)+2 \cos 2 \phi +5\right]-1\right)   N_t\,.
\end{align}

To find the optimal operating point of the interferometer, we return to Eq.~(\ref{eq:sens}). Unlike the information-recycling signal, $\Delta \phi$ does not generally attain a minimum at $\phi = 0$. Here the minimum occurs when $\mathcal{J}(\phi) \equiv V(\jhat_z^2) \cot^2\! \phi + V(\jhat_x^2) \tan^2\! \phi$ is a minimum, which occurs at
\begin{align}
\phi = \tan^{-1}\big[ (V(\jhat_z^2) / V(\jhat_x^2))^{1/4}\big]\,,
\end{align}
and this gives $\min_\phi \mathcal{J} = 2\sqrt{V(\jhat_z^2) V(\jhat_x^2)}$. Consequently, the minimum phase sensitivity is
\begin{align}
	(\Delta \phi_{\min})^2	&= \frac{2\sqrt{V(\jhat_z^2) V(\jhat_x^2)} + C(\jhat_x^2, \jhat_z^2) + \langle   (\jhat_x\jhat_z + \jhat_z\jhat_x)^2 \rangle}{4( \langle  \jhat_z^2 \rangle - \langle  \jhat_x^2 \rangle)^2}\,. \label{app_sens}
\end{align}

If the QST process is a beamsplitter with QST fraction $\mathcal{Q}$, we can use Eqs.~(\ref{J_identities}), (\ref{Jxzzx}), and (\ref{Jxzxz}) to put Eq.~(\ref{app_sens}) in the form
\begin{align}
(\Delta \phi_{\min})^2
= \frac{\mathcal{Q}^2}{\mathcal{F}_b} + \frac{1}{4 \mathcal{Q}^3 \mathcal{F}_b^2}
\Bigg\{&\sqrt{\frac{1}{2}(1-\mathcal{Q})\Big( N_t + \mathcal{Q}(1-\mathcal{Q})
\left[ 6(\mathcal{F}_b - 2 N_t) - N_t^2\right]\Big) \mathcal{A}(\mathcal{Q}, \hat{N}_t)}\notag \\
&+(1-\mathcal{Q})\Big[ \mathcal{Q}^2 \big( 3 \langle   \hat{N}_t^3 \rangle - 2 \mathcal{F}_b N_t \big)
+ 4\mathcal{Q}[4 - \mathcal{Q}(2+\mathcal{Q})(1-\mathcal{Q})] \mathcal{F}_b \notag \\
&-2N_t\big[5 + (1-\mathcal{Q})\big(\mathcal{Q}N_t - 6(1-\mathcal{Q})\big)\big]\Big] \Bigg\}\,,
\label{eq:Deltaphinorecycle}
\end{align}
where
\begin{align}
\mathcal{A}(\mathcal{Q}, \hat{N}_t)
&\equiv 8[\mathcal{Q}\mathcal{F}_b + (1-\mathcal{Q}) N_t]\big[ 10 - 3 \mathcal{Q}(4-\mathcal{Q}) - \mathcal{Q}\big(\mathcal{Q}\mathcal{F}_b + (1-\mathcal{Q}) N_t\big)\big] \notag\\
&\qquad- 24 [3 - \mathcal{Q}(3-\mathcal{Q})]N_t
+ 3 \mathcal{Q}^2 \big[ \mathcal{Q} \langle   \hat{N}_t^4 \rangle + 4(2-\mathcal{Q}) \langle   \hat{N}_t^3 \rangle \big]
\end{align}
and $\mathcal{F}_b = [V(\hat N_t) + N_t (N_t + 2)]/2$ is the quantum Fisher information information for perfect {QST}. Alternatively, we can write the minimum phase sensitivity as
\begin{align}\label{eq:Deltaphinorecycle2}
(\Delta&\phi_{\min})^2\notag\\
&=
\frac{\mathcal{\mathcal{Q}}^4}{\mathcal{F}_a - (1-\mathcal{Q}) N_a} \notag \\
&\quad+\frac{1}{4\left( \mathcal{F}_a - (1-\mathcal{Q}) N_a \right)^2}
\Bigg\{\sqrt{\frac{1}{2}(1-\mathcal{Q})\Big( (1-\mathcal{Q})\left( 6 \mathcal{F}_a - N_a^2\right) - \left( 5 - 6 \mathcal{Q}^2\right) N_a \Big)\tilde{\mathcal{A}}(\mathcal{Q}, \hat{N}_t)}\notag\\
&\quad+(1-\mathcal{\mathcal{Q}})\Big[ 3 \mathcal{\mathcal{Q}}^3 \langle\hat{N}_t^3 \rangle + 4\big(4 - \mathcal{\mathcal{Q}}(2+\mathcal{\mathcal{Q}})(1-\mathcal{\mathcal{Q}})\big) \mathcal{F}_a 
- 2 N_a\big( \mathcal{F}_a + 7 - 2 \mathcal{Q}^4\big)\Big]\Bigg\}\,,
\end{align}
where $N_a = \mathcal{\mathcal{Q}} N_t$, $\mathcal{F}_a = \mathcal{Q}^2 \mathcal{F}_b - (1-\mathcal{Q}) N_a$, and
\begin{align}
\tilde{\mathcal{A}}(\mathcal{Q}, \hat{N}_t) 	
&\equiv \mathcal{Q} \mathcal{A}(\mathcal{Q}, \hat{N}_t) \notag \\
&= 8 \mathcal{F}_a \big( 10 - 3 \mathcal{Q}(4-\mathcal{Q})
-\mathcal{F}_a \big) - 24 \big( 3 - \mathcal{Q}(3-\mathcal{Q}) \big)N_a \notag \\
&\quad+ 3 \mathcal{Q}^3 \big( \mathcal{Q} \langle   \hat{N}_t^4 \rangle + 4 (2-\mathcal{Q}) \langle   \hat{N}_t^3 \rangle \big)\,.
\end{align}
Importantly, when $\mathcal{Q} = 1$, the term in the curly braces on the right-hand side of Eq.~(\ref{eq:Deltaphinorecycle}) or~(\ref{eq:Deltaphinorecycle2}) vanishes, and we have $\Delta \phi_{\min} = 1/\sqrt{\mathcal{F}_b}$, so the phase sensitivity saturates the QCRB, as expected.	

Since $\mathcal{F}_b = \alpha N_t^2 + \mathcal{O}(N_t^2)$ for some positive constant~$\alpha$, we can use the convexity relations $\langle   \hat{N}^4_t \rangle \geq \langle   \hat{N}_t \rangle^4$ and $\langle   \hat{N}_t^3 \rangle \geq \langle   \hat{N}_t \rangle^3$ to determine that, to leading order in $1/N_a$,

\begin{align}
(\Delta \phi_{\min})^2		
&\gtrsim
\frac{\mathcal{\mathcal{Q}}^4}{\alpha N_a^2}
+ (1-\mathcal{Q})\left( \frac{\sqrt{2(6 \alpha - 1)(8 \alpha^2 - 3) + \mathcal{O}(1/N_a)} + 2(3-2\alpha)}{8 \alpha^2 N_a} + \mathcal{O}\left( 1 / N_a^2 \right) \right).
\end{align}
Therefore, for any deviations from perfect QST on the order of $(1-\mathcal{Q})\agt1/N_a$, the Heisenberg scaling is lost, and we return to the standard QNL scaling $\Delta \phi_{\min} \sim \sqrt{(1-\mathcal{Q}) / N_a}$.  Since the initial number of donor particles is typically and desired to be large, in practice Heisenberg scaling is lost for very small departures from perfect {QST}.

We demonstrate this point more concretely in Fig.~2 of the main text, where we plot the phase sensitivity for the specific cases where the donor modes are initially in a twin-Fock state or a two-mode squeezed vacuum state.  For a twin-Fock state, $V(\hat N_t) = 0$, $\langle   \hat{N}_t^3 \rangle =   N_t^3$, and $\langle   \hat{N}_t^4 \rangle =   N_t^4$, and the sensitivity at the optimal operating point is
\begin{subequations}
\begin{align}
(\Delta \phi_\text{TF})^2	
&= \frac{1}{N_a (N_a + 2 \mathcal{Q})^2} \bigg(2 (N_a + 2\mathcal{Q})
+ \sqrt{\tfrac{1}{2}(1-\mathcal{Q})\big[ 1 + 2(1-\mathcal{Q})(N_a - 3 \mathcal{Q})\big]\tilde{\mathcal{A}}_\text{TF}(\mathcal{Q},N_a)} \notag\\
&\hspace{9em}+ 2(1-\mathcal{Q})\big[1 + (N_a - 3\mathcal{Q})(N_a + 2)\big] \bigg) \, , \\
\tilde{\mathcal{A}}_\text{TF}(\mathcal{Q},N_a)	
&= (N_a-2)(N_a+2)(N_a+4)+12(1-\mathcal{Q})\big[2 + N_a(N_a + 3 - \mathcal{Q})\big]\,.
\end{align}
\end{subequations}
For two-mode squeezed vacuum, $V(\hat N_t) =    N_t  (  N_t  + 2)$, $\langle   \hat{N}_t^3 \rangle = 2  N_t ( 2 + 3V(\hat{N}_t))$, and $\langle   \hat{N}_t^4 \rangle = 8 N_t (N_t + 1) (1 + 3V(\hat N_t))$, so the sensitivity at the optimal operating point becomes
\begin{subequations}
\begin{align}
(\Delta \phi_\text{sq})^2	
&=\frac{1}{2 N_a (N_a + 2\mathcal{Q})^2} \bigg(2 (N_a + 2\mathcal{Q})
+ \sqrt{(1-\mathcal{Q})\big[ 1 + 5 (1-\mathcal{Q}) N_a \big]\tilde{\mathcal{A}}_\text{sq}(\mathcal{Q},N_a) \notag}\\
&\hspace{9em}+(1-\mathcal{Q})\Big( 1 + N_a \big[5(1-\mathcal{Q}) + 8(N_a + 2 \mathcal{Q})\big]\Big)\bigg)\,, \\
\tilde{\mathcal{A}}_\text{sq}(\mathcal{Q},N_a)	
&=(1-\mathcal{Q})\big[1-10\mathcal{Q}(1-\mathcal{Q})\big]
+(N_a + 2 \mathcal{Q})\big[9 - 5 \mathcal{Q}(2-\mathcal{Q}) + 8 N_a (N_a + 2)\big]\,.
\end{align}
\end{subequations}

\subsection{Phase sensitivity for the QST process when using information recycling}

If we use the technique of information recycling, the signal we are interested in is $ \hat{\mathcal{S}} = (\jhat_z + \hat L_z)^2 $.  Just as without recycling, there is no interference between sectors with different numbers of particles; the desired moments are thus averages over $p_N=\rho_{NN}$.  Moreover, the anticorrelation of $\jhat_z$ and $\lhat_z$, expressed by $(\jhat_z+\lhat_z)\hat\rho(t_1)=0=-\hat\rho(t_1)(\jhat_z+\lhat_z)$, allows us to convert $\lhat_z$ in these moments to $\jhat_z$.  The anticorrelation, as before, also implies that $\hat\rho(t_1)$ is invariant under rotations about the $z$ axis; in particular, a rotation by $\pi$, which takes $\jhat_x$ to $-\jhat_x$, implies that all terms with an odd number of $\jhat_x$ operators have vanishing expectation value.  In converting $\lhat_z$ to $\jhat_z$, we introduce $\jhat_y$ into our expressions, so this last rule becomes that the only nonvanishing moments are those for which the total power of $\jhat_x$ and $\jhat_y$ is even.

The mean and second moment of the signal are
\begin{align}
\langle\hat{\mathcal{S}}\rangle &= \langle  \jhat_z^2\rangle (\cos \phi-1)^2 + \langle  \jhat_x^2\rangle \sin^2\! \phi\,, \label{S_j} \\
\langle\hat{\mathcal{S}}^2\rangle &= \langle  \jhat_z^4\rangle (\cos \phi-1)^4 + \langle  \jhat_x^4\rangle \sin^4\! \phi + \langle  \jhat_z^2 \jhat_x^2 +\jhat_x^2 \jhat_z^2+ 4  \jhat_z \jhat_x \jhat_x \jhat_z\rangle \sin^2\!\phi (\cos\phi-1)^2\notag\\
&\qquad+  i \langle   (\jhat_z \jhat_x \jhat_y - \jhat_y \jhat_x \jhat_z)\rangle 2 \sin^2\!\phi \cos\phi (\cos\phi-1)+ \langle   \jhat_y^2\rangle \cos^2\!\phi \sin^2\!\phi\,. \label{S_sq_j}
\end{align}
Using \eq{S_j} and \eq{S_sq_j} in $(\Delta \phi)^2 = V(\hat{\mathcal{S}}) / (\partial_\phi \langle \hat{\mathcal{S}}\rangle )^2$ and taking the limit as $\phi\rightarrow 0$ gives
\begin{equation}
(\Delta \phi_{\rm min})^2 = \frac{\langle \hat{J}_y^2\rangle}{4 \langle \hat{J}_x^2\rangle^2} \, .
\end{equation}
Noting that $\langle \hat{J}_y^2\rangle = \langle \hat{J}_x^2\rangle$, we recover $\Delta \phi_{\rm min} = \frac{1}{2\langle \hat{J}_x^2\rangle^{1/2}}$.

Written in terms of conditional expectation values, Eqs.~(\ref{S_j}) and~(\ref{S_sq_j}) become
\begin{align}
\langle\hat{\mathcal{S}}\rangle
=\sum_{N=0}^\infty p_N\bigg(&\frac{1}{2} \sin^2\!\phi \langle   \hat{N}_1\rangle_N
+\frac{1}{2} (\cos\phi-1)^2 \langle   \hat{N}_1^2\rangle_N
+\frac{1}{2} (\sin^2\!\phi -(\cos\phi-1)^2) \langle   \hat{N}_1\rangle_N\langle   \hat{N}_1\rangle_N \bigg)\,,\\
\langle\hat{\mathcal{S}}^2 \rangle
=\sum_{N=0}^\infty p_N\Bigg(&\left(\frac{1}{2} \cos^2\!\phi \sin^2\!\phi-\frac{\sin^4\!\phi}{4}\right)
\langle   \hat{N}_1\rangle_N
+\left((1-\cos\phi)\cos\phi \sin^2\!\phi+\frac{3\sin^4\!\phi}{8}\right)\langle\hat{N}_1^2\rangle_N \notag\\
&+ \frac{3}{4}(\cos\phi-1)^2 \sin^2\!\phi \langle   \hat{N}_1^3\rangle_N
+\frac{1}{8}(\cos\phi-1)^4 \langle   \hat{N}_1^4\rangle_N \notag \\
&+\left(-\sin^2\!\phi \cos\phi+ \frac{6}{4} \cos^2\!\phi \sin^2\!\phi-\frac{\sin^4\!}{4}\right)
\langle   \hat{N}_1\rangle_N \langle   \hat{N}_1\rangle_N \notag\\
&+\left(-\frac{3}{4} (\cos\phi-1)^2 \sin^2\!\phi + \frac{3 \sin^4\!\phi}{4}\right)
\langle   \hat{N}_1^2\rangle_N \langle   \hat{N}_1\rangle_N \notag\\
&+\left(-\frac{1}{2} (\cos\phi-1)^4+ \frac{3}{2} (\cos\phi-1)^2 \sin^2\!\phi\right)
\langle   \hat{N}_1^3\rangle_N\langle   \hat{N}_1\rangle_N \notag\\
&+\left(\frac{3 (\cos\phi-1)^4}{8}-\frac{3}{2} (\cos\phi-1)^2 \sin^2\!\phi+ \frac{3 \sin^4\!\phi}{8}\right) \langle   \hat{N}_1^2\rangle_N\langle   \hat{N}_1^2\rangle_N \Bigg)\,,
\end{align}
Specifying the QST process as a beamsplitter, we again use the identities in the previous section to get
\begin{align}
\langle\hat{\mathcal{S}}\rangle
&= \frac{1}{4} \sin ^2\!\phi\bigg( N_t \mathcal{Q}+\frac{ \langle  \hat{N}_t^2 \rangle\mathcal{Q}^2}{2}\bigg)
-\frac{1}{4}  N_t (\mathcal{Q}-1) \mathcal{Q} (\cos \phi -1)^2\,,\\
\langle\hat{\mathcal{S}}^2\rangle&=
\sin ^4\!\phi  \bigg(\frac{1}{16}  N_t \left(\mathcal{Q}-3 \mathcal{Q}^2\right)+\frac{1}{32}  \langle  \hat{N}_t^2 \rangle \left(3 \mathcal{Q}^2-12 \mathcal{Q}+10\right)
   \mathcal{Q}^2-\frac{3}{32}  \langle  \hat{N}_t^3 \rangle (\mathcal{Q}-2) \mathcal{Q}^3+\frac{3  \langle  \hat{N}_t^4 \rangle \mathcal{Q}^4}{128}\bigg) \notag\\
   &\quad+6 \sin ^2\!\phi  (\cos \phi -1)^2
   \bigg(\frac{1}{16}  N_t \left(2 \mathcal{Q}^2-3 \mathcal{Q}+1\right) \mathcal{Q}+\frac{1}{16}  \langle  \hat{N}_t^2 \rangle (\mathcal{Q}-1)^2 \mathcal{Q}^2-\frac{1}{32}  \langle  \hat{N}_t^3 \rangle
   (\mathcal{Q}-1) \mathcal{Q}^3\bigg) \notag\\
   &\quad+(\cos \phi-1)^4 \bigg(\frac{3}{16}  \langle  \hat{N}_t^2 \rangle (\mathcal{Q}-1)^2 \mathcal{Q}^2-\frac{1}{16}  N_t (\mathcal{Q}-1) \mathcal{Q} \left(6
   \mathcal{Q}^2-6 \mathcal{Q}+1\right)\bigg)\notag\\
   &\quad+\frac{1}{4} \sin ^2\!\phi  \cos ^2\!\phi  \bigg( N_t \mathcal{Q}+\frac{ \langle  \hat{N}_t^2 \rangle
   \mathcal{Q}^2}{2}\bigg)+\frac{1}{2}  N_t (\mathcal{Q}-1) \mathcal{Q} \sin ^2\!\phi  \cos \phi  (\cos \phi -1)\,.
\end{align}
As mentioned in the main text, the optimal operating point is $\phi=0$. To see this we, can expand $\langle   \hat{\mathcal{S}} \rangle$, $\langle   \hat{\mathcal{S}}^2 \rangle$ and $|\partial_{\phi}\langle   \hat{\mathcal{S}} \rangle|^2$  around $\phi=0$.  As only even powers of $\phi$ remain, we have an extremal point here; plotting the sensitivity confirms that $\phi=0$ is indeed a minimum, given by

\begin{align}
(\Delta \phi_{\rm min})^2
=\frac{1}{4 \langle  \jhat_x^2\rangle}
=\frac{2}{\mathcal{Q}^2\langle\hat{N}_t^2\rangle+2\mathcal{Q}N_t}
=\frac{2}{\mathcal{Q}^2V(\hat{N}_t)+N_a(N_a+2)}
=\frac{1}{\mathcal{F}_a}\,.
\end{align}
\end{widetext}

\end{document}